\documentclass[aps,prx,reprint,groupedaddress,amsmath,amssymb,superscriptaddress]{revtex4-2}
\usepackage{graphicx}
\usepackage{float}
\usepackage{multirow}
\usepackage{natbib,twoopt}
\usepackage[breaklinks=true]{hyperref}
\usepackage{xcolor}
\usepackage{amssymb}
\usepackage{stackengine,graphicx}
\usepackage{amsmath}
\usepackage{hyperref}
\usepackage{color}
\usepackage{stackengine}
\usepackage{subfigure}
\usepackage{verbatim}
\usepackage{stmaryrd} 

\newcommand{\rot}{\mathop{\rm rot}\nolimits}
\newcommand{\divv}{\mathop{\rm div}\nolimits}
\newcommand{\df}[2]{\frac{\partial #1}{\partial #2}}

\newcommand{\eps}{\varepsilon}
\newcommand{\om}{\omega}
\newcommand{\kap}{\varkappa}
\newcommand{\bas}{{\bf e}_z}


\begin{document}

\title{Emergent axion response in multilayered metamaterials}

\author{Leon Shaposhnikov}
\affiliation{School of Physics and Engineering, ITMO University, Saint  Petersburg, Russia}

\author{Maxim Mazanov}
\affiliation{School of Physics and Engineering, ITMO University, Saint  Petersburg, Russia}

\author{Daniel A. Bobylev}
\affiliation{School of Physics and Engineering, ITMO University, Saint  Petersburg, Russia}

\author{Frank Wilczek}
\affiliation{Center for Theoretical Physics, Massachusetts Institute of Technology, Cambridge, MA, USA}
\affiliation{Department of Physics, Stockholm University, Stockholm, Sweden}
\affiliation{Department of Physics, Arizona State University, Tempe, AZ, USA}

\author{Maxim A. Gorlach}
\email{m.gorlach@metalab.ifmo.ru}
\affiliation{School of Physics and Engineering, ITMO University, Saint  Petersburg, Russia}

\begin{abstract}
 We consider the design of metamaterials whose behavior embodies the equations of axion electrodynamics.  We derive an effective medium description of an assembly of  magneto-optical layers with out-of-plane magnetization analytically and show how to achieve  effective axion response with tunable parameters.  We display some key predictions and validate them  numerically. 
\end{abstract}

\maketitle

\section{Introduction}

The composition of dark matter is a major open question in physics and cosmology \cite{Bertone2018}. Because presently we have only upper limits on non-gravitational interactions of dark matter, while gravitation is a universal force, several kinds of particles could explain the origin of the dark matter~\cite{Feng2010}.   Axions~\cite{Wilczek78,Weinberg78} are among the most intriguing possibilities, because their existence is  suggested on independent grounds and their predicted properties follow from deep conceptual principles. 

An axion is anticipated to be a light particle, with masses in the range from $\mu$eV to meV favored~\cite{Klaer2017,Buschmann2020}, though smaller masses are also possible. The predicted high phase-space density of cosmological axions allows them to be described  by a  classical pseudoscalar axion field. The electromagnetic coupling of this field leads to additional terms in the Maxwell equations, in the form~\cite{Sikivie1983,Wilczek87}:
\begin{align}
&\rot {\left(\mu^{-1}\,\bf B\right)}=\frac{1}{c}\,\df{(\eps {\bf E})}{t}+\frac{4\pi}{c}\,{\bf j}\notag\\
&\mspace{90mu}+\kap\,\left[\nabla \mathfrak{a}\times{\bf E}\right]+\frac{\kap}{c}\,\df{\mathfrak{a}}{t}\,{\bf B}\:,\label{eq:Axion1}\\
&\divv \left({\eps\,\bf E}\right)=4\pi\rho-\kap\,\left(\nabla \mathfrak{a}\cdot {\bf B}\right)\:,\label{eq:Axion2}\\
&\rot {\bf E}=-\frac{1}{c}\,\df{{\bf B}}{t}\:,\mspace{10mu}\divv {\bf B}=0\:.\label{eq:Axion3}
\end{align}
Here, $\rho$ and ${\bf j}$ are the conventional charges and currents, $\eps$ and $\mu$ are permittivity and permeability of the background medium, and $\kap$ is the axion coupling constant to the electromagnetic field. 

If cosmic axions exist, their coupling constant $\kap$ is extremely feeble, which makes their experimental observation challenging. On the other hand, Eqs.~\eqref{eq:Axion1}-\eqref{eq:Axion3} can be recast as Maxwell's equations in a medium
\begin{align}
&\rot {\bf H}=\frac{1}{c}\,\df{{\bf D}}{t}+\frac{4\pi}{c}\,{\bf j}\:,\label{eq:ED1}\\
&\divv {\bf D}=4\pi\rho\:,\label{eq:ED2}\\
&\rot {\bf E}=-\frac{1}{c}\,\df{{\bf B}}{t}\:,\mspace{10mu}\divv {\bf B}=0\:.\label{eq:ED3}
\end{align}
where the constitutive relations take the form:
\begin{gather}
    {\bf D}=\eps\,{\bf E}+\chi\,{\bf B}\:,\label{eq:Material1}\\
    {\bf H}=-\chi\,{\bf E}+\mu^{-1}\,{\bf B}\:.\label{eq:Material2}
\end{gather}
and $\chi=\varkappa\,\mathfrak{a}$. Hence, if some material is described by the constitutive relations Eqs.~\eqref{eq:Material1}-\eqref{eq:Material2}, its electromagnetic properties are precisely captured by the equations of axion electrodynamics.  

Collective excitations that couple like $\mathfrak a(x,t)$, are known as emergent axions~\cite{Qi2008,Essin2009,Nenno2020,Sekine2021}.    
More common are materials that support a non-trivial constant value of $\mathfrak{a}$. Then the new phenomena arise primarily at interfaces and boundaries.  

In condensed matter physics these sorts of constitutive relations occur in magneto-electrics and multiferroics~\cite{Eerenstein2006,Pyatakov2012}. Such materials were predicted theoretically~\cite{Dzyaloshinskii} and later found in nature, Cr$_2$O$_3$ being the first example followed later by a large class of other magneto-electric materials~\cite{Eerenstein2006,Pyatakov2012}. Multiferroics have received a significant attention from condensed matter community, being of practical use, for instance, for E-Ink technology in electronic books. However, the effective axion response of such structures is relatively weak ($\chi\approx 10^{-3}\div 10^{-2}$) and requires low temperatures in some cases~\cite{Pyatakov2012}. Strong 3D topological insulators also feature effective axion response with quantization of the coefficient~\cite{Nenno2020}.

Parallel investigations occurred in the macroscopic electromagnetism community.  There materials described by the constitutive relations Eqs.~\eqref{eq:Material1}, \eqref{eq:Material2} are known as {\it Tellegen media}, while $\chi$ is termed sometimes Tellegen coefficient. Such materials were first considered by B.D.H.~Tellegen, who suggested a medium consisting of electric and magnetic dipoles attached to each other~\cite{Tellegen}. Tellegen media as well as a wider class of bianisotropic materials were actively investigated~\cite{Kong1974,Tretyakov} and the examples of meta-atoms featuring effective Tellegen response were put forward~\cite{Tretyakov2003,Asadchy2019}.  

{\it Metamaterials\/} are artificially structured media with subwavelength periodicity and engineered, often unconventional electromagnetic properties~\cite{Veselago,Elef,Caloz,Marques,Capolino}. The area of metamaterials has demonstrated such spectacular developments as negative refraction~\cite{Smith2000}, subwavelength imaging~\cite{Pendry2000}, invisibility cloaking~\cite{Leonhardt2006,Pendry2006}.  Recently, it has been proposed that wire metamaterials~\cite{Pendry1998,Belov2003,Simovski2012}  can be useful for cosmic axion detection~\cite{Lawson2019}.

In this Article we bring these strands together, introducing the concept of {\it axion metamaterials\/}.  We demonstrate theoretically that multilayered structures~\cite{Brekh,Yeh,Zhukovsky2015} composed of the conventional magneto-optical materials, should provide practical axion metamaterials, in the sense that they will obey the equations of axion electrodynamics to a good approximation over a substantial range of conditions.  Notably, here the axion response is a design parameter subject to flexible control; in particular, it need not be small.  Qualitative understanding of the comparatively simple effective equations guides us to some distinctive physical predictions, which we have validated quantitatively through numerical simulation of the full dynamics. 

The remainder of this article is organized as follows. In Section~\ref{sec:Design}, we discuss the suggested design of our axion metamaterial. Section~\ref{sec:Calculation} continues with the calculation of the effective axion response $\chi$ for the designed structure, revealing some unexpected aspects of metamaterial homogenization theory. In Section~\ref{sec:Spatial} we examine spatial gradients of $\chi$ and obtain the equations of axion electrodynamics. Next, Sec.~\ref{sec:Optimization} discusses the ways to tailor and control the effective axion response of the designed metamaterial.  In Sec.~\ref{sec:Validation} we validate our effective medium treatment by examining electromagnetic fields in the designed multilayered structure and comparing the results to those predicted by the effective medium model. Finally, we conclude our analysis by discussing the results and outlining future perspectives in Sec.~\ref{sec:Discussion}.

\section{Design of axion metamaterial and symmetry requirements}~\label{sec:Design}

First we examine the symmetry properties of the desired constitutive relations Eqs.~\eqref{eq:Material1},\eqref{eq:Material2}. Since ${\bf E}$ and ${\bf B}$ have different parities under spatial inversion $\mathcal{P}$, effective axion response $\chi$ is odd under inversion, i.e. pseudoscalar. Due to the different behavior of ${\bf E}$ and ${\bf B}$ under time reversal $\mathcal{T}$, $\chi$ is also $\mathcal{T}$-odd. However, it remains invariant under the combined $\mathcal{PT}$ transformation. Such behavior is fully consistent with that expected for an axion field. In electromagnetic context, such response requires external fields breaking the reciprocity of the material.

Furthermore, the constitutive relations Eqs.~\eqref{eq:Material1},\eqref{eq:Material2} have a continuous rotational symmetry. To reconcile that with the fabrication capabilities, we require at least full rotational symmetry of the structure with respect to one axis, $Oz$.

\begin{figure}[t]
	\centering
	\includegraphics[width=0.5\textwidth]{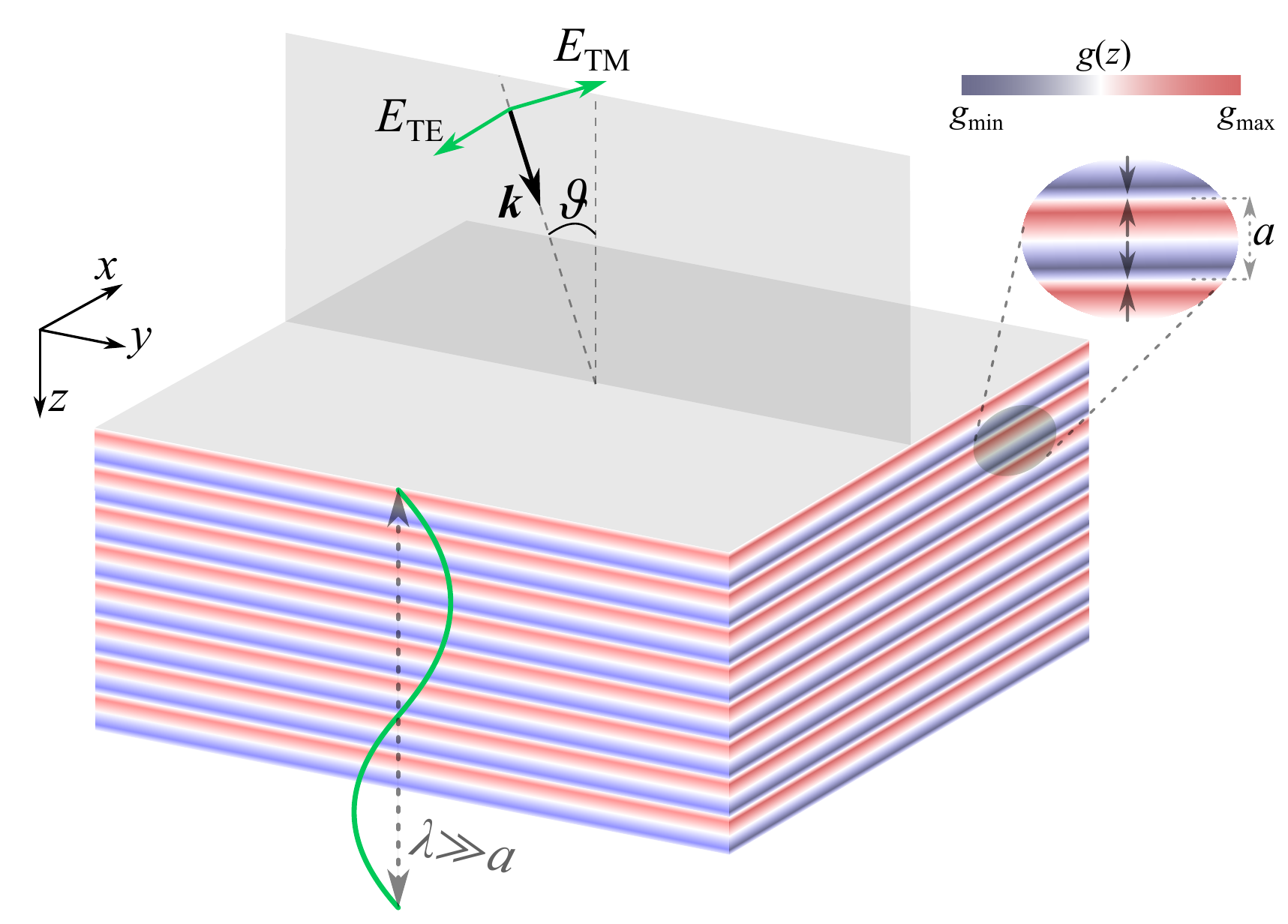}
	\caption{Schematic of the designed multilayered structure composed of gyrotropic layers with spatially varying out-of-plane magnetization schematically shown by the black arrows. The average magnetization of the structure vanishes.    
	}
	\label{fig:Structure}
\end{figure}

As a simple structure fulfilling the above requirements we choose a multilayered metamaterial with out-of-plane magnetization of the layers parallel to $Oz$ axis (Fig.~\ref{fig:Structure}). To exclude the conventional magneto-optical effects such as Faraday effect, we require average magnetization to be zero. The permittivity of a single layer is given by the expression
\begin{equation}\label{eq:permittivity}
    \hat{\eps}=\begin{pmatrix}
        \eps & i\,g(z) & 0\\
        -i\,g(z) & \eps & 0\\
        0 & 0 & \eps
    \end{pmatrix}\:,
\end{equation}
where $g(z)$ is a periodic function with the period $a$, so that its  Fourier expansion
\begin{equation}\label{eq:Gexp}
    g(z)=\sum\limits_{n\not=0} g_n\,e^{inbz}\:,
\end{equation}
$b=2\pi/a$ is the reciprocal lattice period and $g_0\equiv 0$ due to the vanishing average magnetization. From the symmetry point of view, the designed structure breaks time-reversal symmetry. However, the combination of spatial inversion and time reversal leaves it invariant and there is a continuous rotational symmetry with respect to $z$ axis. Hence, the designed structure satisfies the necessary symmetry requirements.

It should be noted that the magneto-electric response of antiferromagnetic structures is known in condensed matter physics for a long time~\cite{Dzyaloshinskii,Pyatakov2012}. However, different magneto-electrics feature a variety of electromagnetic phenomena which distinguish them one from another. A simple picture capturing their electrodynamics is currently lacking, while the ways to control and reconfigure their electromagnetic properties remain practically uncharted. To fill this gap, we investigate electrodynamics of the designed structure in detail, calculate explicitly the effective axion response and show the ways to control its magnitude.

\section{Effective axion response of metamaterial} \label{sec:Calculation}

The effective description of composite media relies on averaging of rapidly oscillating local fields which yields slowly varying {\it macroscopic fields}. The coefficients that relate the macroscopic polarization and magnetization to the averaged fields are associated with the effective material parameters~\cite{Agranovich}. The validity of the effective medium description is largely determined by the period-to-wavelength ratio, $\xi=a/\lambda$, which is considered to be small. Currently, metamaterial homogenization strategy is well established~\cite{Silv2007,Alu2011} with a history of application to the various types of metamaterials~\cite{Belov2003,Silv2007L} including the multilayered ones~\cite{Chebykin2011,Rizza2017,Gorlach2020}. Quite importantly, all homogenization approaches derive the effective material parameters from the bulk properties.

The current problem, however, has a subtle feature. Equations~\eqref{eq:Axion1}-\eqref{eq:Axion3} suggest that the effective axion field which is constant in time and space does not modify Maxwell's equations. Therefore, analysis of the bulk properties in such setting does not allow one to extract the Tellegen coefficient $\chi$. Below, we prove that the analysis of boundary conditions allows one to solve this problem yielding eventually the equations of axion electrodynamics. 

Using the periodicity of the structure, the fields in the metamaterial can be presented in the form:
\begin{equation}\label{eq:Floquetexp}
    \begin{pmatrix}
    {\bf E}({\bf r})\\
    {\bf D}({\bf r})\\
    {\bf B}({\bf r})
    \end{pmatrix}
    =
    \sum\limits_{n=-\infty}^{\infty}\,
    \begin{pmatrix}
    {\bf E}_n({\bf r})\\
    {\bf D}_n({\bf r})\\
    {\bf B}_n({\bf r})    
    \end{pmatrix}
    \,e^{i{\bf k}^{(n)}\cdot{\bf r}}\:,
\end{equation}
where monochromatic $e^{-i\om\,t}$ time dependence is suppressed throughout, ${\bf k}^{(n)}=\left(k_x,0,k_z+nb\right)$, $k_z$ and $k_x$ are the components of Bloch wave vector normal and parallel to the layers, respectively, and $k_y$ can always be set to zero by the choice of the coordinate system. Here, ${\bf E}_0$, ${\bf D}_0$ and ${\bf B}_0$ are the respective averaged (or macroscopic) fields, while ${\bf E}_n$, ${\bf D}_n$ and ${\bf B}_n$ with nonzero $n$ are the amplitudes of the rapidly oscillating Floquet harmonics. In the analysis below, our goal is to derive the equations for the macroscopic fields excluding all rapidly oscillating terms.

The microscopic (non-averaged) fields in the structure satisfy the conventional Maxwell's equations which can be recast in the form
\begin{align}
    &\nabla\left(\divv{\bf E}\right)-\Delta {\bf E}=q^2\,{\bf D}+\frac{4\pi i\,q}{c}\,{\bf j}\:,\label{eq:MainEq1}\\
    &\divv{\bf D}=0\:,\label{eq:MainEq2}
\end{align}
where $q=\omega/c$, ${\bf j}={\bf j}_0\,e^{ik_x x+ik_z z}$ are the external distributed currents exciting the structure. The electric displacement for gyrotropic layers takes the form
\begin{equation}\label{eq:MainEq3}
    {\bf D}=\eps\,{\bf E}-ig(z)\,\left[\bas\times{\bf E}\right]\:.
\end{equation}
Substituting Floquet expansions for the fields [Eq.~\eqref{eq:Floquetexp}] and Fourier series for gyrotropy [Eq.~\eqref{eq:Gexp}] to Eqs.~\eqref{eq:MainEq1}-\eqref{eq:MainEq3}, we recover the set of linear equations that relate the amplitudes of the Floquet harmonics. Applying the perturbation theory in small parameter $\xi=a/\lambda=q/b$, we derive the expressions for the Floquet harmonics of electric field with $n\not=0$ (see Appendix A):
\begin{align}
    &E_{nx}=\frac{ig_n}{n^2\,b^2}\,\left(q^2-\frac{k_x^2}{\eps}\right)\,E_{0y}+O(\xi^3)\:,\label{eq:Enx}\\
    &E_{ny}=-ig_n\,\frac{q^2}{n^2\,b^2}\,E_{0x}+O(\xi^3)\:,\label{eq:Eny}\\
    &E_{nz}=-ig_n\,\frac{k_x}{\eps\,n\,b}\,\left(1-\frac{k_z}{nb}\right)\,E_{0y}+O(\xi^3)\:.\label{eq:Enz}
\end{align}
Using the equation ${\bf B}=-i/q\,\rot{\bf E}$, we can also evaluate the respective Floquet harmonics of magnetic field as ${\bf B}_n={\bf k}^{(n)}\times{\bf E}_n/q$. This yields
\begin{align}
    &B_{nx}=ig_n\,\frac{q}{nb}\,E_{0x}+O(\xi^2)\:,\\
    &B_{ny}=ig_n\,\frac{q}{nb}\,E_{0y}+O(\xi^2)\:,\\
    &B_{nz}=-ig_n\,\frac{k_x\,q}{n^2\,b^2}\,E_{0x}+O(\xi^3)\:,\\
    &B_{0z}=\frac{k_x\,E_{0y}}{q}\:.
\end{align}

Having the explicit expressions for ${\bf E}_n$ and ${\bf B}_n$, we analyze now the boundary conditions at the interface of the metamaterial with air. Clearly, the microscopic fields satisfy the conventional continuity conditions at the interface:
\begin{align}
&\left.{\bf E}_{t}\right|_{z=0}={\bf E}_t^{{\rm out}}\:,\label{eq:MCond1}\\
&\left.B_{z}\right|_{z=0}=B_z^{{\rm out}}\:,\label{eq:MCond2}\\
&\left.{\bf B}_{t}\right|_{z=0}={\bf B}_t^{{\rm out}}\:,\label{eq:MCond3}\\
&\left.\eps\,E_{z}\right|_{z=0}=E_z^{\rm out}\:.\label{eq:MCond4}
\end{align}
However, the microscopic field at the boundary of a metamaterial $z=0$ $\left.{\bf E}\right|_{z=0}=\sum\limits_{n} {\bf E}_n$ is generally different from the averaged field ${\bf E}_0$ due to the contribution of higher-order Floquet harmonics. Keeping the terms up to the first power in small parameter $\xi$ and using the  expressions for the Floquet harmonics above, we recover the following set of boundary conditions for the averaged fields:
\begin{align}
&{\bf E}_{0t}={\bf E}_t^{{\rm out}}\:,\mspace{8mu} &{\bf B}_{0t}-{\bf B}_t^{{\rm out}}=\chi\,{\bf E}_{0t}\:,\label{eq:AxionBC1}\\
&B_{0z}=B_z^{{\rm out}}\:,\mspace{8mu} &\eps\,E_{0z}-E_z^{\rm out}=-\chi\,B_{0z}\:.\label{eq:AxionBC2}
\end{align}

Interestingly, we observe that the tangential components of ${\bf B}_0$ and normal components of $\eps\,{\bf E}_0$ feature the discontinuity. As we prove below, these jumps of the averaged fields at the boundary are a signature of Tellegen medium and the coefficient $\chi$ quantifies the strength of the effective axion response:
\begin{equation}\label{eq:TellegenCoeff}
    \chi=-\frac{iq}{b}\,\sum\limits_{n\not=0}\,\frac{g_n}{n}\:.
\end{equation}

It should be emphasized that the outlined picture of the effective Tellegen medium is valid once the metamaterial is subwavelength ($\xi=a/\lambda\ll 1$) and the effects of the order of $\xi^2$ can be neglected. Counterintuitively, the effective Tellegen response is fully isotropic even though the designed metamaterial has a single axis of continuous rotational symmetry and other rotational axes are lacking. Therefore, the developed description of metamaterial holds for the arbitrary incidence angles. However, as we discuss in Appendix~B, the anisotropy in the electromagnetic response of our metamaterial arises in the second order in period-to-wavelength ratio $\xi$.

Another interesting feature of our system is the dependence of $\chi$ on the structure termination. Indeed, if the boundary is shifted by $\Delta$, the Fourier harmonics of gyrotropy $g(z)$ change from $g_n$ to $g_n\,e^{inb\Delta}$. In the general case, this alters the effective axion response Eq.~\eqref{eq:TellegenCoeff}. This feature is in stark contrast with the behavior of the conventional material parameters which are normally derived from the bulk properties and do not depend on the structure termination~\cite{Silv2007,Alu2011,Gorlach2020}.

\section{Spatial gradients of effective axion response}\label{sec:Spatial}

In the analysis above, we assumed that the metamaterial is periodic and time-independent, which ensures constant $\chi$. To demonstrate the link with axion electrodynamics and prove that $\chi$ is indeed the effective axion response, we generalize our treatment to the case of $\chi$ slowly varying is space. For our metamaterial, this can be achieved by breaking strict periodicity of the structure. To investigate this scenario, we divide the system into blocks with a characteristic size $L$ much larger than the period of metamaterial $a$, but smaller than the characteristic scale of $\chi$ variation.

\begin{figure}[b]
	\centering
	\includegraphics[width=0.5\textwidth]{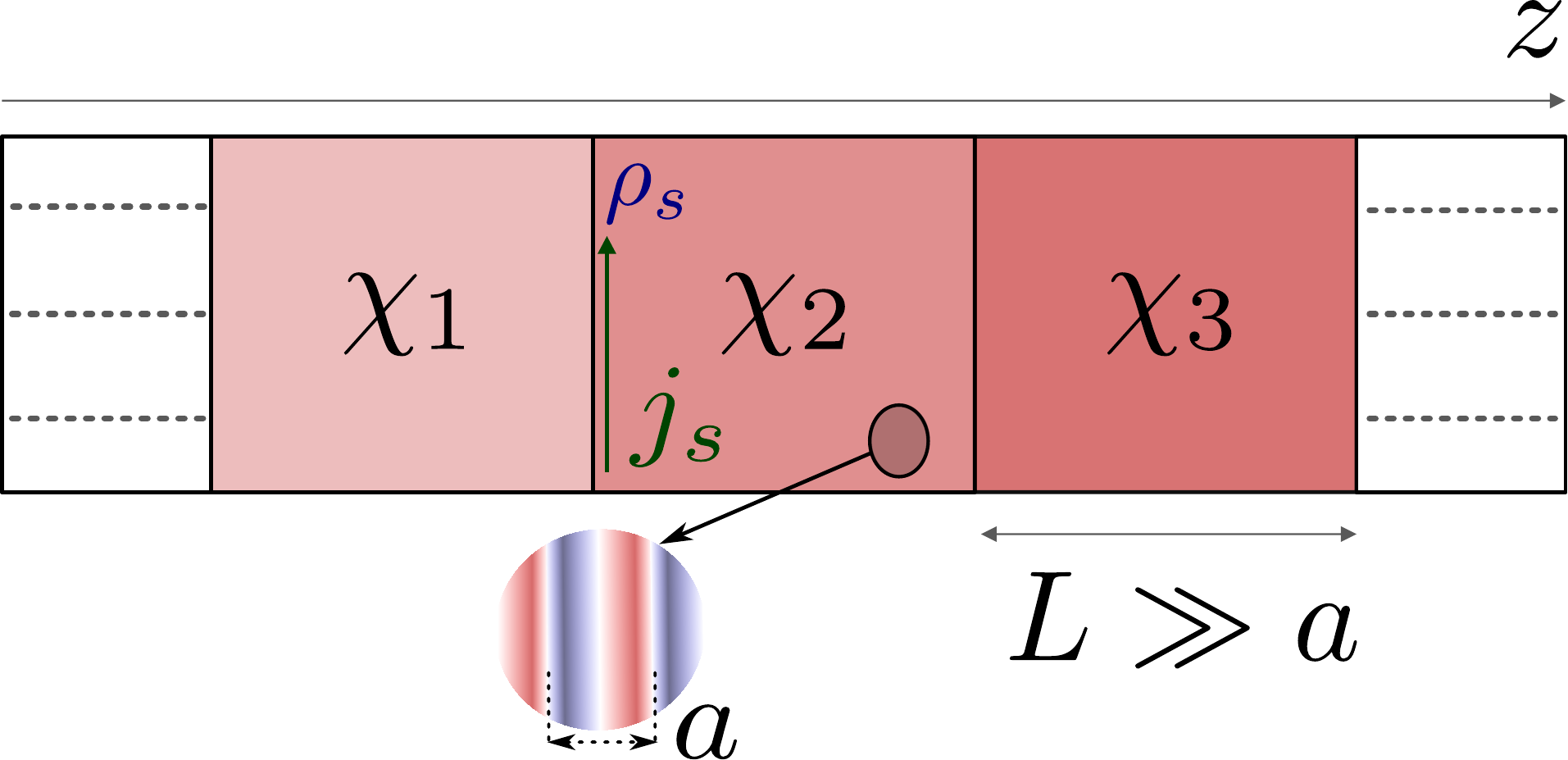}
	\caption{ The sketch of the metamaterial with broken strict periodicity, that is constructed of blocks with close values of $\chi$. The boundaries between the blocks host surface charge with density $\rho_s$ and surface current with density $j_s$.
	}
	\label{fig:grad chi}
\end{figure}

Applying Eqs.~\eqref{eq:AxionBC1},\eqref{eq:AxionBC2} to the boundary of the two adjacent blocks with $\chi_1$ and $\chi_2$ Tellegen coefficients, we recover the following discontinuities in ${\bf B}_t$ and $\eps\,E_z$:
\begin{align}
 &{\bf B}_{2t}-{\bf B}_{1t}=\left(\chi_2-\chi_1\right)\,{\bf E}_t\:,\notag\\
 &\eps\,E_{2z}-\eps\,E_{1z}=-\left(\chi_2-\chi_1\right)\,B_z\:.\notag
\end{align}
On the other hand, such discontinuities result in the surface currents ${\bf j}_s$ and charges $\rho_s$ induced at the boundary between the blocks:
\begin{align*}
    &\frac{4\pi}{c}\,{\bf j}_s=\bas\times\left[{\bf B}_2-{\bf B}_1\right]=\left(\chi_2-\chi_1\right)\,\left[\bas\times{\bf E}\right]\:,\\
    &4\pi\,\rho_s=\eps\,E_{2z}-\eps\,E_{1z}=-\left(\chi_2-\chi_1\right)\,B_z\:.
\end{align*}
To average the obtained distribution of the sources over the scales of the order of $L$, we make the replacement ${\bf j}_s/L\rightarrow {\bf j}$, $\rho_s/L\rightarrow \rho$, $\left(\chi_2-\chi_1\right)\bas/L\rightarrow \nabla\chi$, where ${\bf j}$ and $\rho$ are the respective bulk currents and charges induced due to the gradient of the effective axion response. This procedure yields:
\begin{align}
    \frac{4\pi}{c}\,{\bf j}=\left[\nabla\chi\times{\bf E}\right]\:,\\
    4\pi\rho=-\nabla\chi\cdot{\bf B}\:.
\end{align}
Inserting the obtained expressions into Maxwell's equations with sources, we obtain
\begin{align}
     &\rot{\bf B}=\frac{1}{c}\,\frac{\partial}{\partial t}\left(\eps\,{\bf E}\right)+\left[\nabla\chi\times{\bf E}\right]\:,\label{eq:MetamaterialA1}\\
     &\divv{\left(\eps\,{\bf E}\right)}=-\nabla\chi\cdot{\bf B}\:,\label{eq:MetamaterialA2}\\
     &\rot{{\bf E}}=-\frac{1}{c}\,\df{{\bf B}}{t}\:,\mspace{10mu} \divv{{\bf B}}=0\:.\label{eq:MetamaterialA3}
\end{align}
In the case of time-independent $\chi$, this matches the equations of axion electrodynamics Eqs.~\eqref{eq:Axion1}-\eqref{eq:Axion3} which allows us to interpret $\chi$ as an effective axion response.

\section{Tailoring the effective axion response} \label{sec:Optimization}

A unique advantage of metamaterial platform is the possibility to tailor the effective axion response $\chi$ on demand by manipulating the distribution of magnetization and associated gyrotropy $g(z)$. As discussed in Sec.~\ref{sec:Calculation}, counter-intuitive but technically straightforward way to modify axion response is to change the termination of the metamaterial. This potentially allows not only to change the magnitude, but also to swap the sign of $\chi$.

\begin{figure}[b!]
	\centering
	\includegraphics[width=0.5\textwidth]{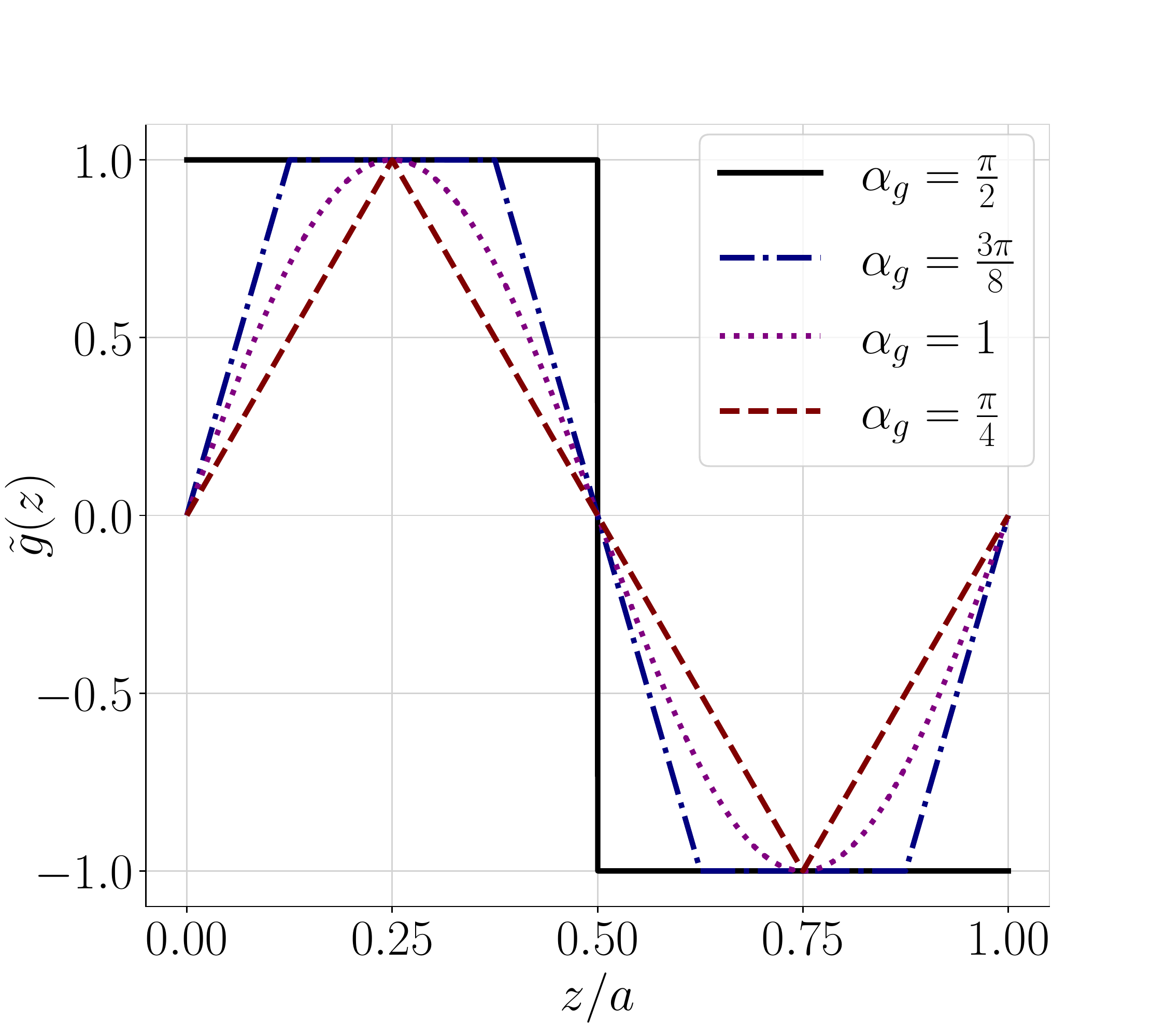}
	\caption{Different spatial distributions of gyrotropy $\tilde{g}(z)$ within the unit cell giving rise to the distinct values of the effective axion response in metamaterial $\chi\propto \alpha_g$.
	}
	\label{fig:magnetic distribution}
\end{figure}

Yet another approach is to tailor the spatial dependence of $g(z)$. To illustrate the dependence of effective axion response on the functional form of $g(z)$, we recast the expression for $\chi$ Eq.~\eqref{eq:TellegenCoeff} in the form
\begin{equation}
    \chi=\alpha_g\,M\,\frac{a}{\lambda}\:,
\end{equation}
where $M=\text{max}\,g(z)$ is the maximal gyrotropy within the unit cell, $\xi=a/\lambda=q/b$ is the period-to-wavelength ratio, while $\alpha_g$ is the dimensionless coefficient that depends on the specific form of $g(z)$ function:
\begin{equation}\label{eq:Tellegen Alpha}
    \alpha_g=\frac{1}{a}\int_0^a(\pi-bz)\tilde{g}(z)dz\:,
\end{equation}
where $\tilde{g}(z)=g(z)/M$ (see the derivation in Appendix C). We examine several representative scenarios of the magnetization distribution within the unit cell, Fig.~\ref{fig:magnetic distribution}, with the same value of the maximal gyrotropy $M$. In each case, we evaluate the dimensionless $\alpha_g$ factor which quantifies the relative strength of the effective axion response at a given period-to-wavelength ratio $a/\lambda$. Comparing the step-function with several other representative examples including the harmonic magnetization modulation, we observe that the stepwise gyrotropy distribution maximizes the strength of the effective axion response.

To elaborate more on this observation, we explicitly derive an upper bond on $\alpha_g$ for the fixed $M$ and $a/\lambda$.
\begin{equation}\label{eq:top est alpha_g}
    |\alpha_g|=\left|\frac{1}{a}\int_0^a(\pi-bz)\tilde{g}(z)dz\right|\leq\frac{1}{a}\int_0^a\left |(\pi-bz)\right |dz=\frac{\pi}{2}\:,
\end{equation}
where we use the fact that $|\tilde{g}(z)|\leq 1$. Since $|(\pi-bz)|=(\pi-bz)\,\text{sgn}\left(\pi-bz\right)$, the upper limit $\alpha_g^{\rm{max}}=\pi/2$ is achieved when $\tilde{g}=\text{sgn}\left(\pi-bz\right)$, which is exactly the step-function profile. In such case, the maximal value of the effective axion response reads
\begin{equation}\label{eq:AxionMax}
    \chi^{\rm{max}}=\frac{\pi}{2}\,M\,\frac{a}{\lambda}\:.
\end{equation}

\section{Validation of the effective medium description}\label{sec:Validation}

After deriving the effective medium picture of the designed metamaterial, we validate this approximate description. For that purpose, we simulate the behavior of axion metamaterial with the stepwise gyrotropy distribution using transfer matrix approach~\cite{Soukoulis2008Apr,Zak1990Sep,Born1999Oct} or full wave numerical techniques that take the full account of the metamaterial microstructure. The obtained results are compared to the predictions of the effective medium model with the effective axion response given by Eq.~\eqref{eq:AxionMax}.

\begin{figure}[h!]
    \centering
	\includegraphics[width=0.45\textwidth]{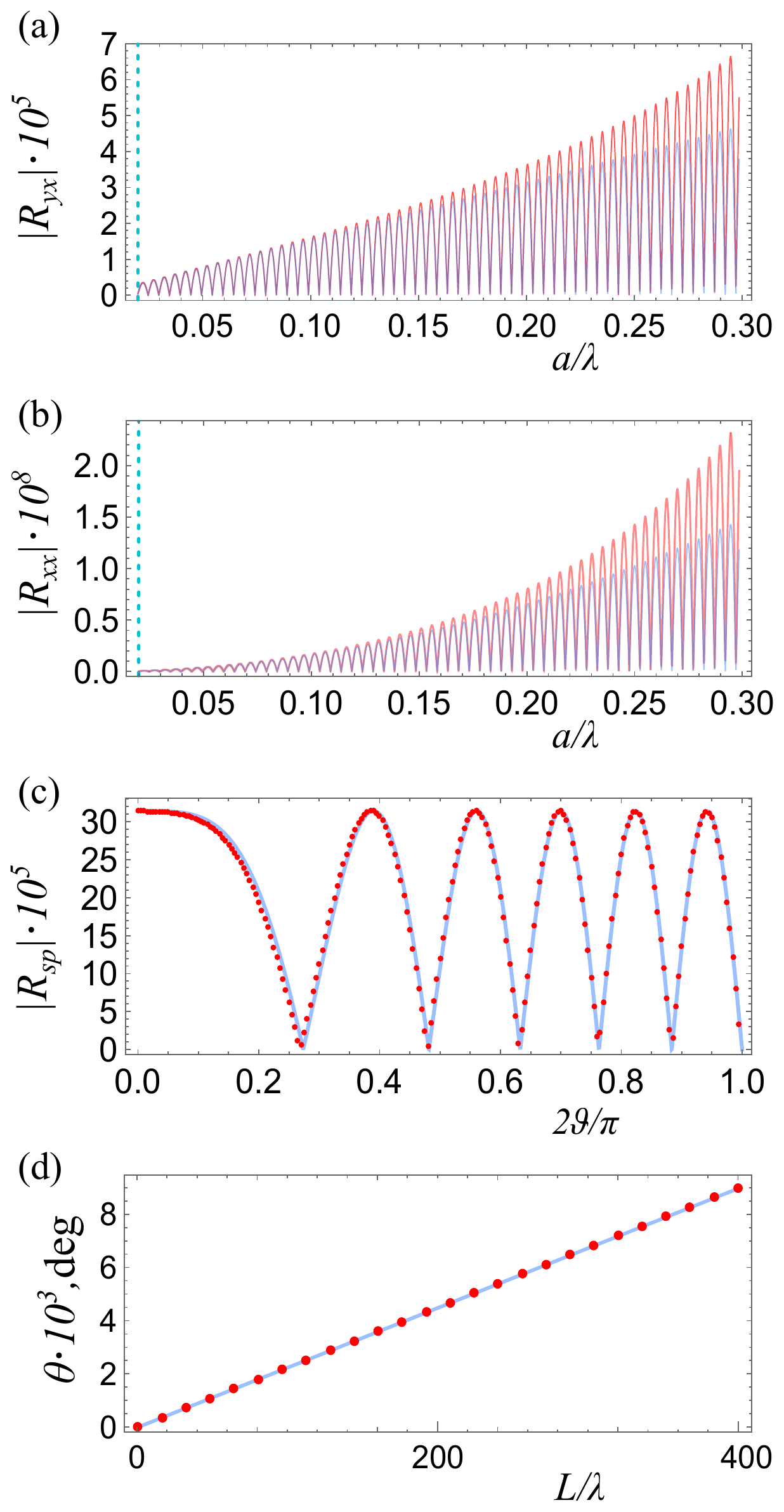}
	\caption{Validation of the effective medium picture for the designed axion metamaterial with the stepwise gyrotropy distribution. Red lines or dots show the results of transfer matrix method, light blue lines are the results of effective medium approximation. (a)~Cross-polarized and (b)~co-polarized reflection coefficients $|r_{yx}|$ and $|r_{xx}|$ for the slab versus the 
    period-to-wavelength ratio $a/\lambda$. Vertical cyan dotted lines mark the ratio $a/\lambda=1/50$ used in the calculations at panels (c)-(d). 
	(c)~Cross-polarized reflection coefficients $|R_{sp}|=|R_{ps}|$ for the slab of axion metamaterial versus incidence angle $\vartheta$. 
	(d)~Averaged rotation angle of the microscopic field near the output facet of axion metamaterial slab with constant gradient of $\chi$. Parameters of the simulations: $\varepsilon_0=\varepsilon=1$; 
 (a)-(b)~$L=100 a$, $g=0.01$; (c)~
 $g=0.01$, $L = 2.75 \lambda$; (d)~
 $g_{\text{max}}=0.01 L/(400 \lambda)$. 
	}
\label{f12}
\end{figure}

First, we examine the reflection of the plane wave at normal incidence from the finite slab of axion metamaterial. Evaluating the effective axion response from Eq.~\eqref{eq:AxionMax} and employing the analytical theory of Tellegen media~\cite{Lindell_1994}, we derive the electric field of the reflected wave ${\bf E}^{\rm r}=\hat{r}\,{\bf E}^{\rm in}$ with $2\times 2$ matrix $\hat{r}$ having the components
\begin{gather}
r_{xx}=r_{yy}=
-
\frac{\left(\chi^2 + \eps - \eps_0\right) \sin\tilde{L}}{\left(\chi ^2+\eps +\eps _0\right) \sin \tilde{L} + 2 i \sqrt{\eps  \eps _0} \cos \tilde{L}}
, \label{Rxx_exact}\\
r_{xy}=-r_{yx}
=
\frac{2 \chi  \sqrt{\eps_0} \sin \tilde{L}}{\left(\chi ^2+\eps +\eps _0\right) \sin \tilde{L} + 2 i \sqrt{\eps  \eps _0} \cos \tilde{L}}
, \label{Rxy_exact}
\end{gather}
where $\tilde{L} = 2\pi\,\sqrt{\eps} L / \lambda_0 = 2 \sqrt{\eps} \pi N a / \lambda_0$ is the optical path inside the slab, $L$ is the thickness of the slab, $N$ is the number of periods in the structure, $\lambda_0$ is the wavelength in vacuum, $\eps$ and $\eps_0$ are the permittivities of the Tellegen medium and host material, respectively. Thus, the polarization plane of the reflected light is rotated, and the reflected light contains both co-polarized and cross-polarized components proportional to $r_{xx}$ and $r_{yx}$, respectively. 

As expected, the major contribution to the co-polarized reflectance comes from the difference between $\eps$ and $\eps_0$. Therefore, to isolate the contributions stemming from the effective axion response $\chi$, we compare the results of transfer matrix method to the analytical expressions Eqs.~\eqref{Rxx_exact}, \eqref{Rxy_exact} for the scenario $\eps=\eps_0$ both for cross-polarized [Fig.~\ref{f12}(a)] and co-polarized [Fig.~\ref{f12}(b)] reflection coefficients. Assuming fixed thickness of the slab $L=100\,a$, we gradually change the frequency of the incident wave thus varying period-to-wavelength ratio $\xi=a/\lambda$. If the metamaterial unit cell is deeply subwavelength ($\xi<0.15$), the two approaches perfectly agree with a typical discrepancy between them of the order of few percents. However, further increase of $\xi$ results in significant errors reaching $50 \%$ for $\xi=0.3$ that make effective medium treatment inadequate.

To further check the validity of our model for plane wave propagation, we fix a sufficiently small period-to-wavelength ratio $\xi=0.02$ and analyze the scenario with nonzero incidence angle $\vartheta$. Figure~\ref{f12}(c) compares the results calculated using the transfer matrix method~\cite{Zak1990Sep} and those obtained from the effective medium approach. In the latter case, we evaluate the transfer matrix for the entire slab employing the relevant boundary conditions and using the fact that the eigenmodes in the Tellegen medium are degenerate and have the refractive index $n = \sqrt{\varepsilon \mu}$. Interestingly, we observe perfect agreement between the two approaches even for the large incidence angles approaching $\pi/2$. This highlights the isotropic nature of the effective axion response despite our model has only one axis of the continuous rotational symmetry.

Next we verify that the equations of axion electrodynamics Eqs.~\eqref{eq:MetamaterialA1}-\eqref{eq:MetamaterialA3} capture the behavior of our metamaterial once its periodicity is broken and a gradient of the effective axion response is introduced. For simplicity, we examine the case of constant gradient that corresponds to Weyl semimetals~\cite{Grushin2012,Zyuzin2012,Asadchy2022} providing an instance of the so-called Carroll-Field-Jackiw electrodynamics~\cite{Carroll1990Feb}. 

A specific prediction of axion electrodynamics in this case is the rotation of  polarization plane of light~\cite{Harari1992Sep} similarly to the Faraday effect in magneto-optical materials. To verify this, we simulate the scenario of normal incidence introducing a linear gradient of axion response $\chi(z) = \chi_{\text{max}} z/L$, where $L$ is the total thickness of the slab. The metamaterial slab is constructed from $N_b \gg 1$ blocks, 
each comprising $N_l$ 
identical bilayers of subwavelength thickness $a \ll \lambda$, with each bilayer consisting of two layers
featuring the same magnitude but opposite orientations of magnetization. 
The absolute value of magnetization is constant within each block, but linearly changes throughout the blocks, which thus creates an approximation of the linear gradient of $\chi(z)$. 
Chosen parameters ensure that the spatial variation of the effective axion field is smooth. We compute the field in the metamaterial close to the output facet and average the field polarization over the block of $N_l \sim 10^2$ layers to exclude rapid oscillations at subwavelength scale (see Appendix D). Figure~\ref{f12}(d) shows the comparison of transfer-matrix results averaged in the described way with the effective medium picture which suggests polarization rotation $\theta = \chi_{\text{max}} / 2$. The agreement between the two approaches is excellent for the various thicknesses $L$ of the slab which confirms the validity of the effective medium description of inhomogeneous axion metamaterials.

\begin{figure}[t]
    \centering
    \includegraphics{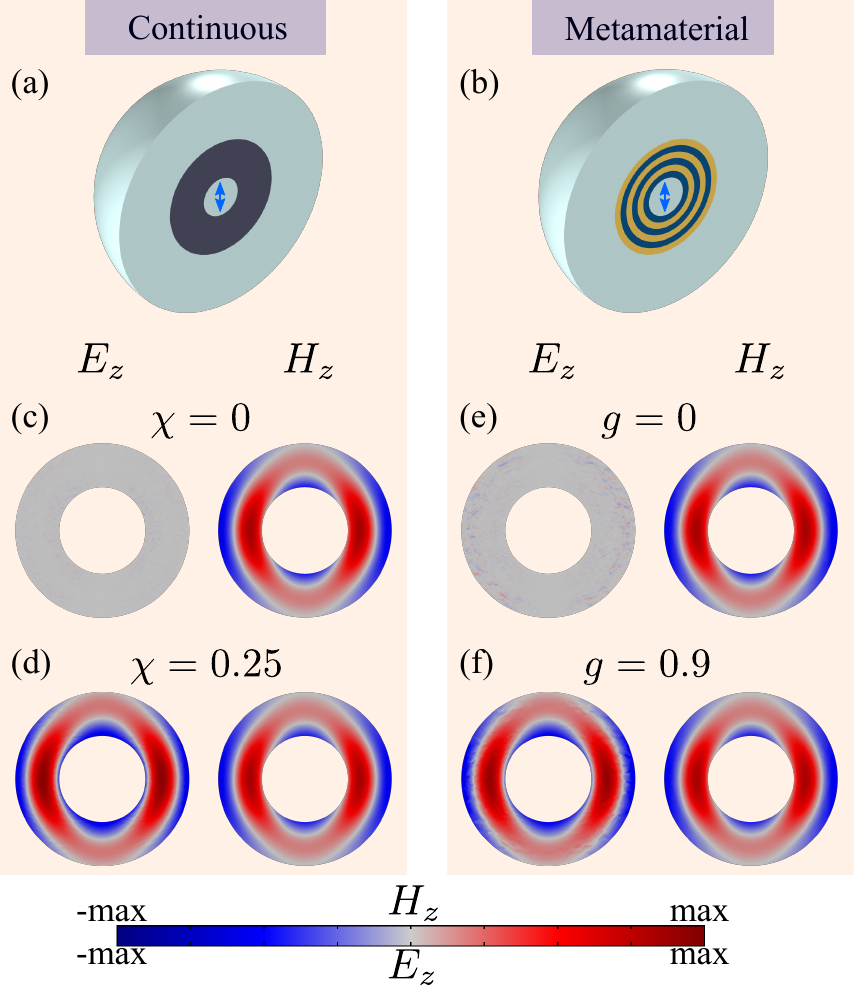}
    \caption{Emergence of electric dipole field for a point magnetic dipole inside the axion spherical shell. (a,b) Sketches of the simulated systems. Panel (a) corresponds to the case of continuous Tellegen medium ($\varepsilon=1$, $\mu=1$), whereas panel (b) shows the metamaterial realization with $\varepsilon=1$, $\mu=1$. The blue arrows represent point magnetic dipole oscillating with the frequency $f=1.0$~GHz. The distance from the dipole to the shell inner boundary is $\lambda/4$, the shell thickness is $\lambda/2$. Period of the layered structure is $a=\lambda/6$. (c-f) Fields corresponding to the continuous medium (c,d) and layered metamaterial (e,f) when the effective axion response is either zero (c,e) or nonzero (d,f).}
\label{fig:simulation}
\end{figure}

Finally, we examine whether the effective medium treatment remains adequate when the metamaterial is excited by the external sources. Specifically, we analyze the fields produced by the oscillating point magnetic dipole surrounded by the axion shell [Fig.~\ref{fig:simulation}(a)]. The theory~\cite{Wilczek87} predicts that the axion shell hybridizes electric and magnetic responses such that the field outside is a superposition of magnetic and electric dipole fields [Fig.~\ref{fig:simulation}(c,d)]. To validate this physics, we simulate the magnetic dipole inside the designed metamaterial [Fig.~\ref{fig:simulation}(b)] and analyze the scenarios with layers magnetization  from zero [Fig.~\ref{fig:simulation}(e)] to some fixed nonzero value [Fig.~\ref{fig:simulation}(f)] matching the magnitude of $\xi$ used in the effective medium calculation. Examining the obtained field patterns [Fig.~\ref{fig:simulation}(f)], we recover that our metamaterial indeed generates an electric dipole field with the induced electric dipole parallel to the magnetic one. Interestingly, the oscillating electric dipole inside the axion shell also induces an effective collinear magnetic moment as further discussed in Appendix~E.


\section{Discussion and conclusions}\label{sec:Discussion}

We have proposed a practical design for axion metamaterial and calculated its effective Tellegen response $\chi$ from the first principles.  While the building blocks of our metamaterial are conventional magneto-optical layers, axion response for the entire structure emerges through their combined action.  

Interestingly, the strength of the axion response depends on the structure termination which in turn affects the boundary conditions.  The importance of boundary terms is well appreciated in the emergent axion theory, but represents an important subtlety in metamaterial homogenization.

Our derivation of the effective axion response complements the existing toolkit of more abstract theoretical methods such as dimensional reduction procedures~\cite{Qi2008} or analysis of quantum field theory anomalies~\cite{Fujikawa1979,Fujikawa1980}.

While the condensed-matter community has been studying magneto-electric and multiferroic materials for a long time, connecting them to metamaterials opens new vistas.  The dual possibilities of producing metamaterials with corresponding properties, but on larger length scales and with tunable properties; or of ``scaling up'' to new kinds of metamaterials using magneto-electric and multiferroic building-blocks, both deserve much further attention, as does the possibility of achieving time-dependent control.  


\section*{Acknowledgments}

We acknowledge Timur Seidov, Dr. Alexander Millar, Prof. Alexandra Kalashnikova and Prof. Pavel Belov for valuable discussions. Theoretical models were supported by Priority 2030 Federal Academic Leadership Program. Numerical simulations were supported by the Russian Science Foundation (Grant No.~20-72-10065). L.S., M.M., D.A.B. and M.A.G. acknowledge partial support by the Foundation for the Advancement of Theoretical Physics and Mathematics ``Basis''.

\section*{Appendix A. Floquet harmonics of electric field}\label{app:Floquet}

This Appendix supplements the discussion in Sec.~\ref{sec:Calculation} and provides the derivation of electric field Floquet harmonics. The starting point of this derivation is the set of Eqs.~\eqref{eq:MainEq1}-\eqref{eq:MainEq3}. Using the Fourier expansion of gyrotropy and Floquet expansion of electromagnetic field,  Eqs.~\eqref{eq:Gexp},\eqref{eq:Floquetexp}, we derive the set of scalar equations for the respective Floquet harmonics:
\begin{gather}
    \left(k_z^{(n)}\right)^2\,E_{nx}-k_x k_z^{(n)}\,E_{nz}=q^2\,D_{nx}\:,\label{eq:Aux1}\\
    \left[k_x^2+\left(k_z^{(n)}\right)^2\right]\,E_{ny}=q^2\,D_{ny}\:,\label{eq:Aux2}\\
    -k_x\,k_z^{(n)}\,E_{nx}+k_x^2\,E_{nz}=q^2\,D_{nz}\:,\label{eq:Aux3}\\
    k_x\,D_{nx}+k_z^{(n)}\,D_{nz}=0\:,\label{eq:Aux4}\\
    D_{nx}=\eps\,E_{nx}+i\,\sum\limits_{n'\not=n}\,g_{n-n'}\,E_{n'y}\:,\label{eq:Aux5}\\
    D_{ny}=\eps\,E_{ny}-i\,\sum\limits_{n'\not=n}\,g_{n-n'}\,E_{n'x}\:,\label{eq:Aux6}\\
    D_{nz}=\eps\,E_{nz}\:.\label{eq:Aux7}    
\end{gather}
From this system of equations, we calculate $E_{nx}$, $E_{ny}$ and $E_{nz}$ up to the second power in small parameter $\xi=a/\lambda=q/b$.

Equation~\eqref{eq:Aux2} yields $E_{ny}\approx q^2/(n^2\,b^2)\,D_{ny}+O(\xi^3)$. In turn, Eq.~\eqref{eq:Aux6} suggests $D_{ny}=-ig_n\,E_{0x}+O(\xi^2)$. Combining these two results, we recover
\begin{equation}\label{eq:EnyFl}
    E_{ny}=-ig_n\,\frac{q^2}{n^2\,b^2}\,E_{0x}+O(\xi^3)\:.
\end{equation}

Next, Eq.~\eqref{eq:Aux5} yields $D_{nx}=ig_n\,E_{0y}+O(\xi^2)$ and Eq.~\eqref{eq:Aux4} allows to calculate $D_{nz}$: $D_{nz}=-k_x/k_z^{(n)}\,D_{nx}=-ig_n\,k_x/(nb)\,\left(1-k_z/(nb)\right)\,E_{0y}+O(\xi^3)$. Using Eq.~\eqref{eq:Aux7}, we immediately evaluate
\begin{equation}\label{eq:EnzFl}
    E_{nz}=-ig_n\,\frac{k_x}{\eps\,nb}\,\left(1-\frac{k_z}{nb}\right)\,E_{0y}+O(\xi^3)\:.
\end{equation}
Finally, we use Eq.~\eqref{eq:Aux1} to calculate $E_{nx}$ via already found $D_{nx}$ and $E_{nz}$ which yields
\begin{equation}\label{eq:EnxFl}
    E_{nx}=\frac{ig_n}{n^2\,b^2}\,\left(q^2-\frac{k_x^2}{\eps}\right)\,E_{0y}+O(\xi^3)\:.
\end{equation}
Equations Eqs.~\eqref{eq:EnyFl}-\eqref{eq:EnxFl} define higher-order Floquet harmonics via  averaged fields which allows us to construct the effective description of metamaterial in terms of the averaged fields.

\section*{Appendix B. Effective permittivity of axion metamaterial}\label{app:Eps}

In Section~\ref{sec:Calculation}, we focused our attention on the derivation of the effective axion response $\chi$. For completeness, we discuss here the derivation of the effective permittivity of metamaterial, $\hat{\eps}^{\rm{eff}}$ keeping the terms up to $\xi^2$. We start from Eq.~\eqref{eq:MainEq3} which can be recast in the form
\begin{equation*}
   {\bf D}_0=\eps\,{\bf E}_0+i\,\sum\limits_{n\not=0}\,g_{-n}\,\left[{\bf E}_n\times\bas\right]\:. 
\end{equation*}
Combining this with the expressions for ${\bf E}_{n}$ Floquet harmonics, Eqs.~\eqref{eq:Enx}-\eqref{eq:Enz}, we recover

\begin{equation*}
  \begin{pmatrix}
      D_{0x}\\
      D_{0y}\\
      D_{0z}
  \end{pmatrix}
  =
  \begin{pmatrix}
  \eps^{{\rm eff}}_{xx} & 0 & 0\\
  0 & \eps^{{\rm eff}}_{yy} & 0\\
  0 & 0 & \eps^{{\rm eff}}_{zz}
  \end{pmatrix}\,
  \begin{pmatrix}
      E_{0x}\\
      E_{0y}\\
      E_{0z}
  \end{pmatrix}\:,  
\end{equation*}
where the components of the effective permittivity tensor in the chosen coordinate system ($k_y=0$) calculated with the precision up to $\xi^2$ read:
\begin{gather}
    \eps^{{\rm eff}}_{xx}=\eps+q^2\,\sum\limits_{n\not=0}\,\frac{g_n\,g_{-n}}{n^2\,b^2}\:,\label{eq:Epsxx}\\
    \eps^{{\rm eff}}_{yy}=\eps+\left(q^2-\frac{k_x^2}{\eps}\right)\,\sum\limits_{n\not=0}\,\frac{g_n\,g_{-n}}{n^2\,b^2}\:,\label{eq:Epsyy}\\
    \eps^{{\rm eff}}_{zz}=\eps\:.\label{eq:Epszz}
\end{gather}
We observe that the correction to the effective permittivity is of the order of $\xi^2$, while the effective axion response $\chi$ is stronger, being of the order of $\xi$. Once terms proportional to $\xi^2$ are taken into account, the metamaterial becomes anisotropic and spatial dispersion effects emerge. Note also that $\eps^{{\rm eff}}$ does not depend on the choice of the unit cell as it is typical for the bulk properties of metamaterials~\cite{Gorlach2020}. In addition, the terms proportional to $\xi^2$ modify the boundary conditions.


\section*{Appendix C. Analysis of the formula for the effective axion response}\label{app:FourierSum}

In this Appendix, we rewrite the expression for the effective axion response $\chi$ [Eq.~\eqref{eq:TellegenCoeff}] in the form more convenient for calculations. For that purpose, we transform the sum of the Fourier components $g_n$ as follows
\begin{widetext}
\begin{gather}
-i\,\sum\limits_{n\not=0}\,\frac{g_n}{n}=-i\,\sum\limits_{n\not=0}\,\frac{1}{n\,a}\,\int\limits_{0}^{a}\,g(z)\,e^{-inbz}\,dz=-\frac{i}{a}\,\int\limits_{0}^{a}\,g(z)\,\left(\sum\limits_{n\not=0}\,\frac{e^{-inbz}}{n}\right)\,dz\notag\\
=
-\frac{i}{a}\,\int\limits_{0}^{a}\,g(z)\,\left[\ln\left(1-e^{-ibz}\right)-\ln\left(1-e^{ibz}\right)\right]\,dz=\frac{1}{a}\,\int\limits_0^a\,g(z)\,(\pi-b\,z)\,dz\:.
\end{gather}
\end{widetext}
Hence, the effective axion response can be recast in the form
\begin{equation}
    \chi=\frac{1}{\lambda}\,\int\limits_0^a\,g(z)\,(\pi-b\,z)\,dz\:.
\end{equation}



\section*{Appendix D. Details of the transfer matrix method}\label{app:Eps}

Here, we briefly discuss the calculation of reflection and transmission coefficients, Faraday rotation and ellipticity via the transfer-matrix method at normal incidence, both for the case of isotropic Tellegen media and the designed multilayered metamaterial. To examine oblique incidence, we use the general form of transfer matrices derived in Ref.~\cite{Zak1990Sep}.

First, we discuss the application of the transfer-matrix method to the calculation of transmission/reflection amplitudes under normal incidence in the homogeneous isotropic Tellegen medium described by the constitutive relations Eqs.~\eqref{eq:Material1}-\eqref{eq:Material2}. The relevant amplitudes have been calculated for some special cases of the homogeneous Tellegen media in Refs.~\cite{Lindell_1994,Tretyakov}, though using a different form of the material equations.

As an amplitude column vector, we choose the $4$-component vector of tangential field components and define the transfer matrix $\hat{M}$ as
\begin{equation}
    \begin{pmatrix}
    \textbf{E}(z) \\
    \textbf{e}_z \times \textbf{B}(z)
    \end{pmatrix}
    =
    \hat{M}
    \begin{pmatrix}
    \textbf{E}(0) \\
    \textbf{e}_z \times \textbf{B}(0)
    \end{pmatrix}
. 
\end{equation}
This choice of the basis is particularly convenient in view of implementing the relevant boundary condition, Eq.~\eqref{eq:AxionBC1}, which translates into a boundary transfer-matrix between between two isotropic Tellegen media with Tellegen parameters $\chi_1$ and $\chi_2$: 
\begin{equation}
    \hat{M}_{1\shortrightarrow2}
    = 
    \begin{pmatrix}
    \hat{I} & 0 \\ 
    (\chi_2 - \chi_1) \hat{\textbf{e}}^\times_z & \hat{I}
    \end{pmatrix}
, 
\end{equation}
where $\hat{I}$ is the two-dimensional unit matrix, and the matrix 
\begin{equation}
    \hat{\textbf{e}}^\times_z
    = 
    \begin{pmatrix}
    0 & -1 \\ 
    1 & 0
    \end{pmatrix}
\end{equation}
describes the effect of the vector product $\textbf{e}_z \times$ on the transverse fields. 

Since the waves propagate in the bulk of Tellegen medium identically to their propagation in isotropic dielectric, the transfer-matrix describing the bulk of the Tellegen medium coincides with that of a dielectric with the same permittivity $\varepsilon$ and reads:


\begin{equation}
\hat{M}_d(z)
=
\left(\begin{array}{cc}
\cos (k z) & -i \sin (k z) / \sqrt{\varepsilon} \\
-i \sin (k z) \sqrt{\varepsilon} & \cos (k z)
\end{array}\right)
,
\end{equation}
where $k = \sqrt \varepsilon \omega/c$. Hence, the full transfer-matrix for the free-standing Tellegen slab of length $L$ reads 
\begin{equation}
\label{tmatrix_1}
\hat{M}
=
\hat{M}_{2\shortrightarrow1}
\hat{M}_d(L)
\hat{M}_{1\shortrightarrow2}
. 
\end{equation}
Then, all the necessary reflection and transmission amplitudes could be deduced from the following system: 
\begin{equation}
\label{free-standing-eqns}
\left(\begin{array}{c}
\textbf{E}^t \\
-\sqrt{\varepsilon_0} \textbf{E}^t
\end{array}\right)=\left(\begin{array}{ll}
\hat{M}_{11} & \hat{M}_{12} \\
\hat{M}_{21} & \hat{M}_{22}
\end{array}\right)\left(\begin{array}{c}
\textbf{E}^{in}+\textbf{E}^r \\
-\sqrt{\varepsilon_0} \textbf{E}^{in}+\sqrt{\varepsilon_0} \textbf{E}^r
\end{array}\right)
, 
\end{equation}
where upper indices ${in},t,r$ correspond to incident, transmitted and reflected fields, respectively, and $\varepsilon_0$ is the permittivity of the medium surrounding the Tellegen slab. For example, the reflection matrix 
\begin{eqnarray}
\hat{R}
=
\left(\begin{array}{cc}
r_{xx} & r_{xy} \\
r_{yx} & r_{yy}
\end{array}\right)
\end{eqnarray}
connecting the incident and reflected fields ($\textbf{E}^r=\hat{R} \textbf{E}^{in}$) 
reads
\begin{gather*}
\hat{R}=-\left[\sqrt{\varepsilon_0} \hat{M}_{11} + \hat{M}_{21} + \varepsilon_0 \hat{M}_{12}+\sqrt{\varepsilon_0} \hat{M}_{22}\right]^{-1}\cdot
\\
\left[\sqrt{\varepsilon_0} \hat{M}_{11} + \hat{M}_{21} - \varepsilon_0 \hat{M}_{12}-\sqrt{\varepsilon_0} \hat{M}_{22}\right]
, 
\end{gather*}
which yields Eqs.~\eqref{Rxx_exact}-\eqref{Rxy_exact} in the main text.

Second, we discuss the application of the transfer-matrix method to the designed  multilayered metamaterial. At normal incidence, boundary conditions for TM and TE modes coincide, and  the modes with left and right circular polarizations (LCP and RCP) propagate without any mixing. Accordinly, we employ the exact transfer matrices in the basis of circular polarizations with amplitude vector $( E^{LCP}_{\rightarrow}, E^{LCP}_{\leftarrow}, E^{RCP}_{\rightarrow}, E^{RCP}_{\leftarrow})^T$, where indices $\leftarrow, \rightarrow$ indicate the direction of propagation of the corresponding plane waves: either along $\textbf{e}_z$ or in the opposite direction. The transfer matrix realizing the boundary conditions between the two layers $i,j$ then reads~\cite{Soukoulis2008Apr}  
\begin{eqnarray}
\hat{M}_{ij}
=
\frac{1}{2}
\begin{pmatrix}
1 + \eta^\pm_{ij} & 1 - \eta^\pm_{ij} & 0 & 0 \\
1 - \eta^\pm_{ij} & 1 + \eta^\pm_{ij} & 0 & 0 \\
0 & 0 & 1 + \eta^\mp_{ij} & 1 - \eta^\mp_{ij} \\
0 & 0 & 1 - \eta^\mp_{ij} & 1 + \eta^\mp_{ij}
\end{pmatrix}
, 
\end{eqnarray}
where $\eta^\pm_{ij} = n^\pm_{i} / n^\pm_{j}$, and $n^\pm_i$ is the refractive indices for LCP/RCP modes in the $i^{\text{th}}$ layer, which for layers with positive $H_z$ read $n^\pm_{\rightarrow} = \sqrt{\varepsilon_0 \pm g}$, while for layers with negative $H_z$ read $n^\pm_{\leftarrow} = \sqrt{\varepsilon_0 \mp g}$ (in air, $n^\pm_{air} = 1$). 
To model the perfect mirror, we use the boundary transfer-matrix for the perfect electric conductor: 
\begin{eqnarray}
\hat{M}_{\text{PEC}}
=
\frac{1}{2}
\begin{pmatrix}
1 & 1 & 0 & 0 \\
1 & 1 & 0 & 0 \\
0 & 0 & 1 & 1 \\
0 & 0 & 1 & 1
\end{pmatrix}
.
\end{eqnarray}
Transfer-matrices describing the propagation in the bulk read 
\begin{eqnarray}
\hat{T}_{i}
=
\begin{pmatrix}
e^{I n^+_i P} & 0 & 0 & 0 \\
0 & e^{-I n^+_i P} & 0 & 0 \\
0 & 0 & e^{I n^-_i P} & 0 \\
0 & 0 & 0 & e^{-I n^-_i P}
\end{pmatrix}
, 
\end{eqnarray}
where $P = 2\pi a/\lambda_0$, and $\lambda_0$ is the light wavelength in air. 
The resultant transfer matrix describing the entire multilayer with even number $2N$ of layers then reads
\begin{eqnarray}
\label{TM}
\hat{M}_{\text{AFM}}
=
\hat{M}_{\shortleftarrow\,\text{air}} \hat{M}_{\shortrightarrow \shortleftarrow} (\hat{M}_{\shortleftarrow \shortrightarrow} \hat{T}_{\shortleftarrow} \hat{M}_{\shortrightarrow \shortleftarrow} \hat{T}_{\shortrightarrow})^{N} \hat{M}_{\text{air}\,\shortrightarrow}
, 
\end{eqnarray}
while the transfer-matrix for the same structure backed with perfectly conducting mirror reads 
\begin{eqnarray}
\label{TMPEC}
\hat{M}_{\text{AFM+PEC}}
=
\hat{M}_{\text{PEC}} \hat{M}_{\shortleftarrow\,\text{air}}^{-1} \hat{M}_{\text{AFM}}
. 
\end{eqnarray}
Co- and cross-polarized reflection/transmission coefficients are then calculated as~\cite{Soukoulis2008Apr} 
\begin{equation}
\label{RTransferM}
\begin{split}
&r_{xx}
=
\frac{1}{2} \left(\frac{M_{2,1}}{M_{2,2}}+\frac{M_{4,3}}{M_{4,4}}\right)
, 
r_{yx}
=
\frac{1}{2i} \left(\frac{M_{2,1}}{M_{2,2}}-\frac{M_{4,3}}{M_{4,4}}\right)
, \\
&t_{xx}
=
\frac{1}{2} \left( \frac{\det(\hat{M}_\text{L})}{M_{2,2}} + \frac{\det(\hat{M}_\text{R})}{M_{4,4}}
\right) 
, \\
&t_{yx}
=
\frac{1}{2i} \left( \frac{\det(\hat{M}_\text{L})}{M_{2,2}} - \frac{\det(\hat{M}_\text{R})}{M_{4,4}}
\right) 
, 
\end{split}
\end{equation}
where $\hat{M}_\text{L}$ and $\hat{M}_\text{R}$ represent the upper left and lower right matrix blocks of the resultant transfer matrix, respectively. 
The angle of polarization rotation $\theta^{r,t}$ and ellipticity $\eta^{r,t}$ are calculated via  
\begin{equation}
\label{ethtransferM}
\begin{split}
2\theta^s
=
\tan^{-1}\left(\frac{2 \Re(K_s)}{1-\left| K_s\right| ^2}\right)
, \\
2\eta^s
=
\sin^{-1}\left(\frac{2 \Im(K_s)}{1+\left| K_s\right| ^2}\right) 
, 
\end{split}
\end{equation}
where $K_s = s_{yx}/s_{xx}$, and $s \in \{r,t\}$.

\begin{figure}[ht]
    \centering
	\includegraphics[width=0.35\textwidth]{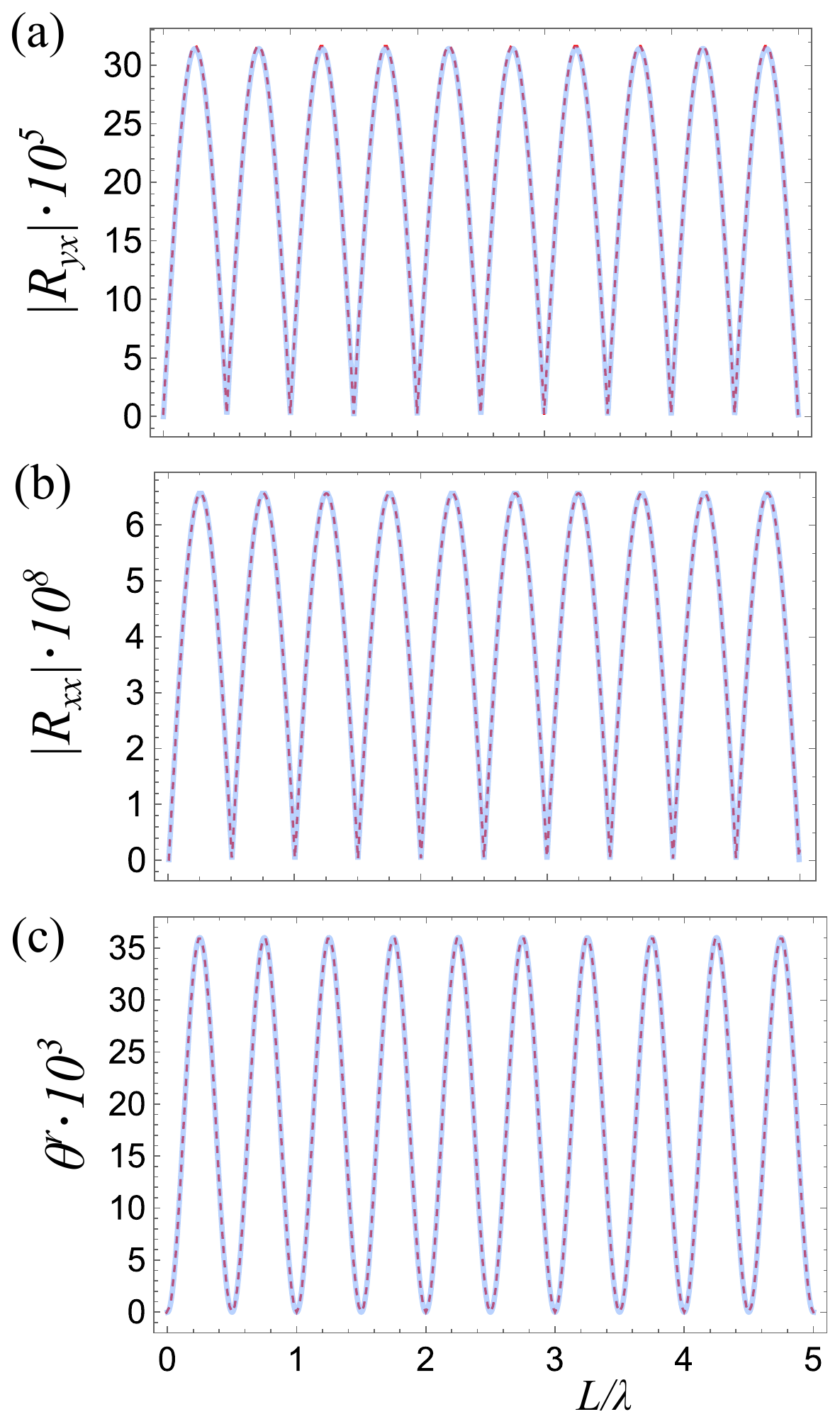}
        \caption{Calculation of the reflection coefficient from the slab of axion metamaterial backed by the ideal mirror. Blue solid and red dashed lines show the prediction of the effective medium approach and transfer matrix method, respectively.
        (a,b) Cross- and co-polarized reflection coefficients versus the thickness $L$ of the slab for the fixed period-to-wavelength ratio $a/\lambda=0.02$. Effective medium results, Eq.~\eqref{Rxy_exact}, include second-order corrections $\propto \chi^2$ to the effective permittivity. 
        (c) Rotation of polarization plane $\theta^r$ for the light reflected from the mirror-coated axion metamaterial slab versus the total slab thickness $L$. Parameters: $a/\lambda=0.02$, $\varepsilon_0=1.0$, $g=0.01$. 
	}
\label{ftheta}
\end{figure}

While the main text discusses the reflection from the free standing slab of axion metamaterial, we also compare numerical and analytical results for the reflection from the axion metamaterial backed by the ideal mirror. For that purpose, we employ the same transfer matrix approach as above, but with a different full transfer-matrix 
\begin{equation}
\label{tmatrix_2}
\hat{M}_{\text{PEC}}
=
\hat{M}_d(L)
\hat{M}_{1\shortrightarrow2}
, 
\end{equation}
while forcing the transmitted field $\textbf{E}^t$ in Eq.~\eqref{free-standing-eqns} to vanish at the PEC boundary. 
This allows us to calculate the reflection amplitudes $r_{xx}$ and $r_{yx}$, and  the rotation angle of the reflected light via Eq.~\eqref{ethtransferM}: 
%
\begin{gather}
\label{theta_exact}
\theta^{r}
=
\frac{1}{2}\,
\arctan\left[
\frac{
4 \chi\psi \sqrt{\varepsilon_0} \sin^2\tilde{L}
}
{
-4 \chi^2 \varepsilon_0 \sin^4\tilde{L}
+\psi^2
}
\right]
, 
\end{gather}
%
where $\psi=\varepsilon  \cos^2\tilde{L} + \sin^2\tilde{L} \left(\varepsilon_0-\chi ^2\right)$ and the ellipticity of the reflected light vanishes. For the typical scenario $\chi \ll 1$, Eq.~\eqref{theta_exact} gives  
\begin{eqnarray}
\theta^{r}
\approx
\frac{2 \chi  \sqrt{\varepsilon_0} \sin^2\tilde{L}}{\varepsilon  \cos^2\tilde{L} + \varepsilon_0 \sin^2\tilde{L}}
. 
\end{eqnarray}
This prediction is reproduced with high precision by the transfer matrix method,  Fig.~\ref{ftheta}.

Note that calculating the predictions of the effective medium model with the precision up to $\chi^2$, we also have to take into account the corrections to the effective permittivity of the metamaterial that also have the order of $\chi^2$ (see Appendix B).

For the multilayered structure under study with $g_n = i g (e^{-i \pi n} -1)/(\pi n)$, this correction reads $\Delta\varepsilon^{\text{eff}} \equiv (\varepsilon_{xx,yy}^{\text{eff}} - \varepsilon) =  (a/\lambda)^2 g^2 (\pi^2/12)$. Using this value of the permittivity $\eps \rightarrow \eps + \Delta\varepsilon^{\text{eff}}$ in Eqs.~\eqref{Rxx_exact}-\eqref{Rxy_exact}, we recover that the calculated cross- and co-polarized reflection coefficients $r_{xy}$, $r_{xx}$ perfectly agree with the results of transfer matrix approach [see Fig.~\ref{f12}(b)].

Finally, in order to calculate the average polarization of the field near the output facet of the multilayer with a linear gradient of effective axion response, we compose such a multilayer out of $N_b \gg 1$ blocks of length $L/N_b$ where $L$ is the total slab thickness, each comprising $N_l \gg L / (\lambda N_b)$ individual bilayers with the same magnitude but opposite orientation of magnetization. In such case, each block can be ascribed an effective Tellegen coefficient which differs from block to block, defining a spatially varying effective axion response $\chi(z)$. 
To calculate Fig~\ref{f12} in the main text, the following parameters are used: $N_b = 100$, $N_l = 200$, $a = \lambda/50$. 

\section*{Appendix E. Point electric dipole inside axion metamaterial}

\begin{figure}[ht]
    \centering
    \includegraphics{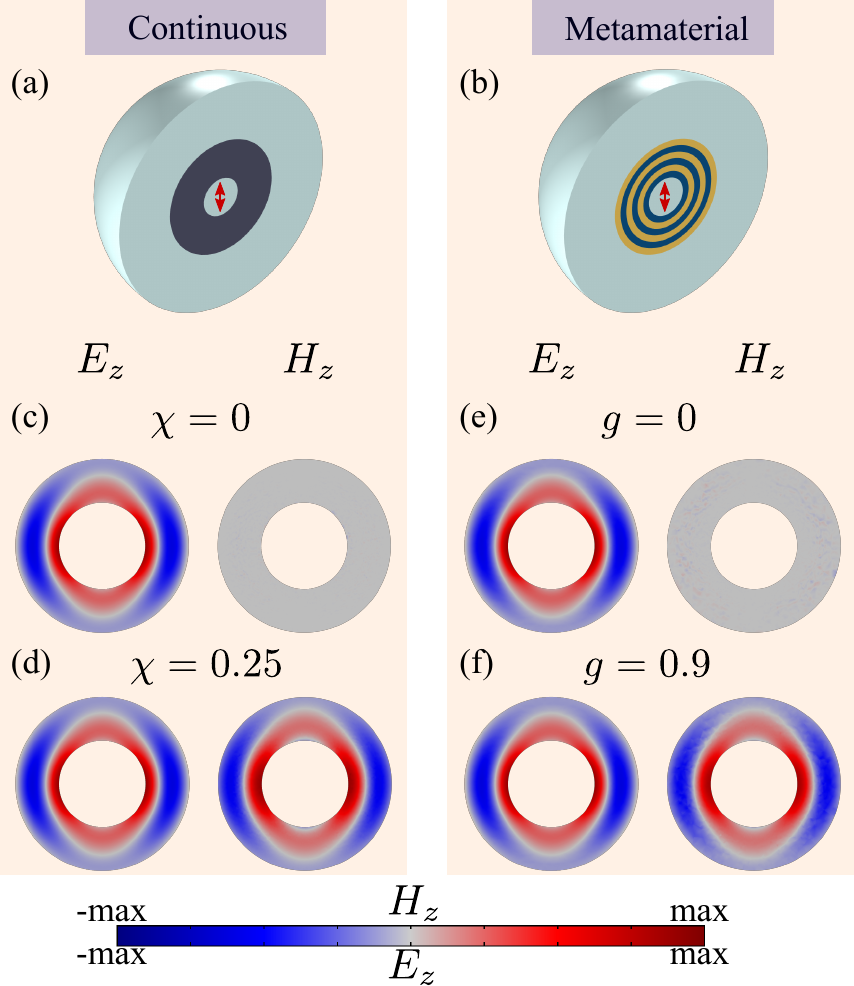}
    \caption{Emergence of magnetic dipole field for a point electric dipole inside an axion spherical shell. (a,b) Sketches of the simulated systems. Panel (a) corresponds to the case of continuous Tellegen medium ($\varepsilon=1$, $\mu=1$), panel (b) shows the metamaterial realization with $\varepsilon=1$, $\mu=1$. The red arrows represent point electric dipole oscillating with the frequency $f=1.0$~GHz. The distance from the dipole to the shell inner boundary is $\lambda/4$, the shell thickness is $\lambda/2$. Period of the layered structure is $a=\lambda/6$. (c-f) Fields corresponding to the continuous medium (c,d) and layered metamaterial (e,f) when the effective axion response is either zero (c,e) or nonzero (d,f).}
\label{fig:simulationapp}
\end{figure}

As discussed in the main text [Fig.~\ref{fig:simulation}], point magnetic dipole surrounded by the axion spherical shell produces the combination of electric and magnetic dipole fields. A similar effect is observed if point electric dipole is placed inside the axion shell [Fig.~\ref{fig:simulationapp}(a,b)]. In this case, the field outside the shell is expected to be a combination of collinear electric and magnetic dipoles [Fig.~\ref{fig:simulation}(d)]. This expectation is confirmed by the full-wave numerical simulations of electric dipole inside the designed metamaterial [Fig.~\ref{fig:simulation}(f)], which highlights once again the validity of the effective medium description.

\bibliography{AxionLib}

\end{document}